# A Review of Quantum Cybersecurity: Threats, Risks and Opportunities

Md Jobair Hossain Faruk*, Sharaban Tahora†, Masrura Tasnim‡, Hossain Shahriar§, Nazmus Sakib¶,

*Department of Software Engineering and Game Development, Kennesaw State University, USA
†Department of Computer Science, Kennesaw State University, USA
‡Institute for Cybersecurity Workforce Development, Kennesaw State University, USA
§Department of Information Technology, Kennesaw State University, USA
¶Department of Computer Science and Engineering, University at Buffalo, The State University of New York, USA

{*mhossa21, †stahora, ‡mtasnim1}@students.kennesaw.edu & {§hshahria}@kennesaw.edu & {¶nsakib2}@buffalo.edu

*Abstract*—The promise of quantum computing is not speeding up conventional computing rather delivering an exponential advantage for certain classes of problems, with profound implications for cybersecurity for instance. With the advent and development of quantum computers, cyberspace security can surely become the most critical problem for the Internet in near future. On contrary, prosaic quantum technology can be promising to transform cybersecurity. This research aims to synthesize basic and fundamental studies concerning quantum cybersecurity that can be emerged both as a threat and solution to critical cybersecurity issues based on a systematic study. We provide a comprehensive, illustrative description of the current state-of-the-art quantum computing and cybersecurity and present the proposed approaches to date. Findings in quantum computing cybersecurity suggest that quantum computing can be adopted for the betterment of cybersecurity threats while it poses the most unexpected threats to cybersecurity. The focus and depth of this systematic survey not only provide quantum and cybersecurity practitioners and researchers with a consolidated body of knowledge about current trends in this area but also underpins a starting point for further research in this field.

*Index Terms*—Quantum Computing, Cybersecurity, Quantum Cybersecurity, Quantum Security, Quantum Threats

## I. INTRODUCTION

With the progress of digitization in all aspects of human life has paved to the storage of all sorts of data in databases [1]. Due to the sensitivity of stored digitalized information, protecting and ensuring the security of data has increased significantly with the advancement of novel technology. Insecure data may lead to losing data to hackers or malware infections can have far greater consequences. The term cybersecurity refers to securing information in general and in recent years, has become indispensable to protect trust and confidence in the digital infrastructure whilst respecting fundamental values like equality, fairness, freedom, and privacy of users or organizations [2], [3]. With the gradual cybersecurity breaches on the rise, communities in academia, companies, and institutions are concerned about protecting digital resources from cybercriminals, hacktivists, and advanced persistence [4].

The field of cybersecurity, as a result, preparing for further advancement by adopting futuristic technologies including AI, Quantum Computing, Blockchain, and Data Science [5], [6]. Quantum computing describes as an emerging domain that adopts the concepts of quantum mechanics and intersects with other domains including mathematics, physics, and computer science to perform computations [7]. This novel technology can address various complex scientific challenges in conventional computing technology and generate opportunities. In the advancement of futuristic technologies, cybersecurity infrastructure will become obsolete in near future [8].

Quantum computing may enable advance opportunities for cybersecuirty practitioners from one perspective and poses risks to the cybersecurity environment at the same time. Now is an opportune time for a fruitful discussion among research community who share an interest in both quantum computing and cybersecurity where emerging quantum computing technology could break most modern cryptography [9]. The main idea of this paper aims to study the possible intersection between Quantum Computing and Cybersecurity. The survey also intends to study the recently proposed approaches to introduce the prospects of quantum computing in cybersecurity from both opportunities and risks perspectives. Our ultimate goal is to spurring discussions and furthering ideas in interdisciplinary research to reflect on Quantum Computing cybersecurity for concerning stakeholders.

*A. Contribution*

In this paper, we intend to study the quantum cybersecurity that has the potential in empowering quantum computing for reducing cyber-related threats. The primary contributions of the paper are as follows:

- We synthesize basic and primary studies concerning quantum cybersecurity that can be emerged both as a threat and solution to critical cybersecurity issues.
- We included 20 primary study related to quantum computing and cybersecurity that can provide suitable benchmarks for comparative analysis against similar research.

- We conduct a comprehensive review of the primarily included papers and present the findings including potential ideas, approaches, and consideration in the fields on quantum computing and cybersecurity.
- We discuss the findings, challenges and provide guidelines for future research from both perspective in this field.

*B. Paper Organization*

The rest of the paper is structured as follows: Section III provides a thorough overview of Quantum Computing and Cybersecurity. Section II explains research design and methodology which this paper were systematically adopted for analysis and provides research questions on both positive and negative aspects of quantum cybersecurity. Section III presents related studies and Section IV discusses findings and future research direction. Finally, Section V concludes the paper.

## II. RESEARCH DESIGN & METHODOLOGY

A systematic literature review [10], [11] has been conducted to find the proper existing study. In this section, we discuss the goal and provide identified research questions. We also provide the overview of research design, inclusion and exclusion criteria, and final papers selection methods that shall pave the study for a thorough evaluation of quantum cybersecurity.

*A. Research Goals*

The advent and development of quantum physics have surely been one of the most unexpected threats to cybersecurity. After careful evaluation of our goal for this paper, we considered the following research questions to be addressed in this study:

- RQ1: What is Quantum Computing and how can it be intersected with Cybersecurity?
- RQ2: What is the potential of quantum cybersecurity as Opportunities and possible Risks?
- RQ3: What are the improvements that shall be carried out in quantum cybersecurity?

*B. Primary Studies Selection*

A "Search Process" was implemented to identify research papers that address our topic of study depicted in Fig. 1 [12]. We first prepared the potential search strings related to the study topic which contained the following keywords:

- "Quantum Cybersecurity"
- "Quantum Computing for Cybersecurity"
- "Cybersecurity for Quantum Computing"
- "Quantum-enabled Cybersecurity" and
- "Post-Quantum Cybersecurity"

We selected various scientific databases and specific search strings were applied during the analysis throughout. The search was carried out with not only the selected keywords but also the title itself on March 01st, 2022. All of the studies published up to the date were considered in this study. The scientific databases that were used for procuring these papers including:

- "Google Scholar"
- "IEEE Xplore Digital Library"
- "ScienceDirect"
- "ACM Digital Library"
- "SpringerLink" and
- "arXiv e-Print Archive"

*C. Search Procedure and Results*

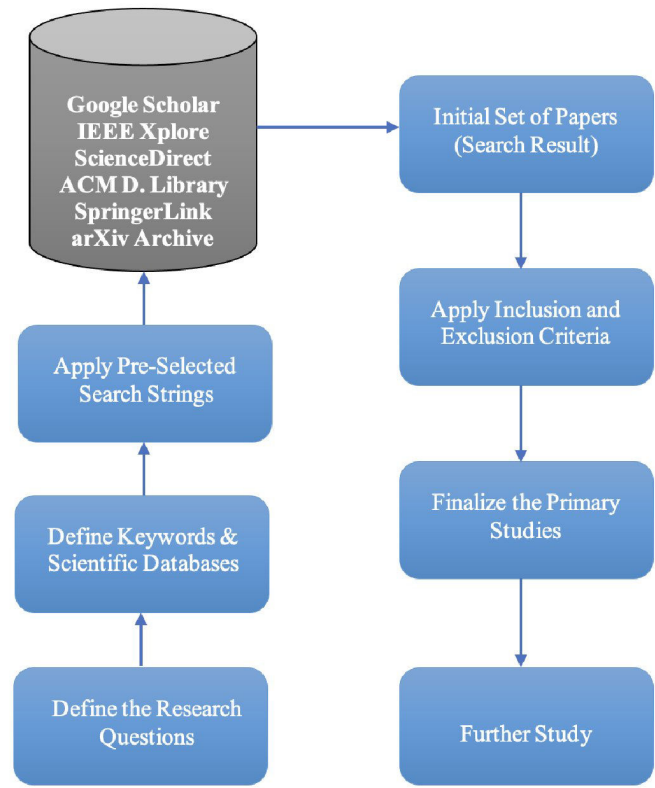

Fig. 1. Attrition of Systematic Literature thorough processing

During the initial search process, we adopt filtration procedure right after applying search strings. Considering time constraint, we chose a specific time-frame that allowed finding research papers published between 2016 and 2022. Furthermore, additional filters were placed in each database to narrow our search of relevant research materials. IEEE Xplore included Conferences and Journals while ScienceDirect required us to select Computer Science as the subject area and research articles for article type. Total 354 research papers were found during the initial search. After completing the initial search process, we applied inclusion and exclusion criteria.

*D. Inclusion and Exclusion Criteria*

An exclusion and inclusion process based on (i) duplicate papers (ii) full-text availability, (iii) peer-reviewed, and (iv)

papers that are not related to the paper topic was conducted to prune off research papers that had aspects that were not related to our literature review as well as duplicates that appeared during the initial search. We later consider various in-depth screening process including empirical methodology and findings in the papers, emerging aspects of quantum cybersecurity, and proposed solution. After the in-depth screening process that accounted for the publication title, abstract, experimental results and conclusions shortened the list to 24 papers for our study. Table II displays the details of the inclusion and exclusion process.

TABLE I
INCLUSION AND EXCLUSION CRITERIA FOR THE PRIMARY STUDIES

| Condition (Inclusion) | Condition (Exclusion) |
| --- | --- |
| The study must be related to Software Engineering Process, Methodology, Blockchain-Oriented Software Development & SPI | Studies focusing on other topics than inclusion papers: blockchain in healthcare, business for instance |
| Papers are not duplicated in different databases | Similar papers in different databases |
| Studies are not duplicated in different scientific databases | Similar Studies in multiple scientific databases |
| Papers contain information related to software process and blockchain technology | Papers that do not cover expected domain |
| Peer-reviewed papers published in a conference proceeding or journal | Non-peer-reviewed papers |
| Studies that are available in the full format & in English | Studies are not available fully and Non-English studies |

TABLE II
GENERALIZED TABLE FOR SEARCH CRITERIA

| Scientific Database | Initial Keyword Search | Total Inclusion |
| --- | --- | --- |
| Google Scholar | 45 | 12 |
| IEEE Xplore | 30 | 08 |
| ScienceDirect | 01 | 00 |
| Springer Link | 05 | 01 |
| ACM | 03 | 01 |
| ResearchGate | 05 | 02 |
| Total | 89 | 24 |

According to Kitchenham and Charters [13], we applied a quality assessment in our primary studies to check the relevance of selected studies with the research questions. We selected five papers randomly for the quality assessment process by following the authors [14].

- **Stage 1: Quantum Computing and Cybersecurity:** The study shall be focused on both quantum computing and cybersecurity where specific issues should be well-described which addresses RQ2.
- **Stage 2: Context:** In order to identify accurate interpretation of the study, we focused on contextual aspects including research objectives and findings.
- **Stage 3: Related Framework:** The papers shall consists of appropriate framework that intersect both quantum computing and cybersecurity which addresses our RQ1 and RQ3.
- **Stage 4: Framework Context:** Presented frameworks shall be related to quantum cybersecurity conceptually or in real-world scenario.
- **Stage 5: Research Findings:** The papers shall have provided sufficient evaluation result or research findings. The studies also expected to provide future research direction due to the emerging development which addresses RQ3.

### E. Publication Over Time

The idea of quantum computing was presented in 1980 and it has been receiving more and more attention during the last few decades [15]. However, ideas of quantum cybersecurity is still in its infancy and thus, we identified papers on this topic since 2016 and illustrate the number of papers pave published. Fig. 2 depicts the evolution of published studies on quantum cybersecurity according to the criteria defined in our study design. Similar to most research areas, conferences generally publish preliminary results while journals publish more mature studies. Academic journals primarily reach an academic audience, while articles also have practitioners as their readers. According to our findings, publication in various journals, conferences, and online articles is an upward trend within the area of quantum cybersecurity.

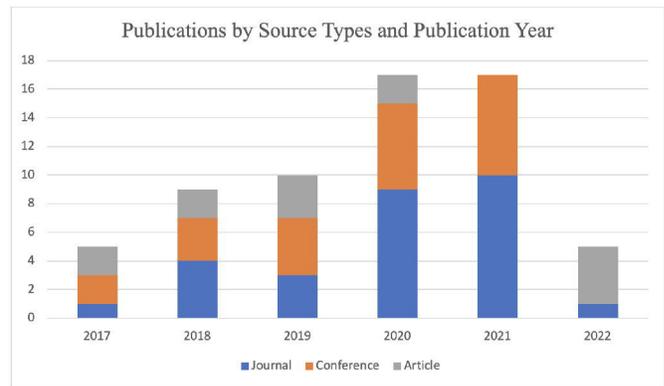

Fig. 2. Publications by Source Types and Publication Year

### F. Significant Keywords Counts

Identifying keywords is crucial for a systematic study that gives proper results against the goal of the study. In this study, we outline a number of keywords that are most common in our selected primary study. Table 4 depicts the keywords and number of times some specific words appeared in all of the primary studies.

TABLE III
SPECIFIC KEYWORDS COUNTS OF THE KEYWORDS IN THE PRIMARY
STUDIES.

| Common Keywords | Total Count |
|---|---|
| Quantum | 6111 |
| Security | 2048 |
| Quantum Computing | 771 |
| Attacks | 979 |
| Cryptography | 957 |
| Quantum Bits or Qubits | 310 |
| Threats | 838 |
| Vulnerability | 834 |
| Post-Quantum | 628 |
| Cybersecurity | 525 |
| Annealing | 522 |
| Quantum Annealing | 389 |
| KeyShield | 283 |
| Quantum-Safe | 260 |
| Traffic Flow | 217 |
| Algebraic Attack | 116 |
| Quantum Resistance | 116 |
| Symmetric Ciphe | 106 |

## III. RELATED STUDIES

- The paper [16] investigates the current state of the art on post-quantum cryptosystems and how can be applied to blockchains and distributed ledger technologies (DLT). Blockchain technology comprises public-key cryptography and hash functions and this is where quantum computing attacks using Grover's and Shor's algorithms can be impactful, poses threats. The rapid advancement of quantum computing has been fueled by attacks that use Grover's and Shor's algorithms, putting hash functions and public-key cryptography at risk. The paper address this issue based on post-quantum schemes and provides extensive comparisons of public-key encryption and digital signature for blockchain. Despite its resilience to quantum threats, the multivariate-based cryptosystem relies on system complexity via solving multivariate equations. Authors introduced five different types of post-quantum cryptosystems and various examples of encryption and digital signature scheme implementations are provided in Fig. 3. Public-key post-quantum cryptosystems comprises of (i) code-based, (ii) multivariate-based, (iii) lattice-based, (iv) supersingular elliptic curver isogeny, and (v) hybrid cryptosystems. In addition, lattice-based cryptosystems comprise a set of dimensional points having a periodic structure. Post-quantum signature techniques guarantee proper authentication and low counterfeiting.

- Althobaiti and Dohler [17] discuss quantum computing and the Internet of Things as an evolving networking paradigm that connects many expedients securely to the internet for cybersecurity protocol. The paper focuses on techniques related to security in a post-quantum IoT and examines the nature of the 3GPP IoT security in the environment of post-quantum. Notably, the researchers discuss features of the fifth-generation (5G) networks, propose an improvement, and how computers on quantum that can conciliate security. The researchers evaluated various methodologies that need to be enforced to protect IoT in the recent and post-quantum realm from vulnerabilities in the contemporary architecture and implementation of IoT. The analysis revealed the current protection of IoT including confidentiality, authentication, and integrity where methods of analyzing the security threat in the IoT architecture are associated with different layers for data levels. The threats established for IoT implementation include malware, direct channel attacks, and brute force attacks. The paper concludes that 5G networks needed more refinement and development due to the disadvantages since 5G has more vulnerable to attacks using quantum computing. To address the problem, researchers introduce promising lattice-driven cryptographic techniques as quantum resistance security models that shall be suitable for mitigating the emerging threats in the current and post-quantum world.

- M. Y. AL-Darwbi [18] focus on implementation of digital signature schemes based on using error-correcting codes that directed to the research of opportunities of implementation of digital signature schemes based on using error-correcting codes. It authorizes cryptographers to make strategy that are unsusceptible to classic cryptanalysis and cryptanalysis which uses quantum computing. The paper illustrate the principles of the classic digital signature scheme CFS using a Niederreiter-like transform, and also propose a new approach that enables an implementation of signature according to the McEliece transformations. Proposed technique provides additional protection against special attacks, and a comparative analysis and characterization of resistance to classic and quantum cryptanalysis, complexity of necessary transformations and length of generated signatures are made.

The researchers extend the research by proposing a scalable and quantum-safe key management scheme called KeyShield [19]. The KeyShield ensures the highest security level and exact solution to an underdetermined linear system of equations with considering several factors including forward/backward secrecy, minimal trust, guessing resistance, quantum resistance, scalability, computational cost, storage overhead, messages overhead, and entity management. By utilizing a linear system of equations as key management scheme is high computation complexity, whereas KeyShiled address the issue with two main solutions, members grouping and using a banded matrix. KeyShield is supported by mathematical analysis and proofs when required. The demonstartion indicates that KeyShiled outperforms state-of-the-art schemes in several aspects, including quantum-resistance, computation cost, message overhead, storage cost, and rekeying delay.

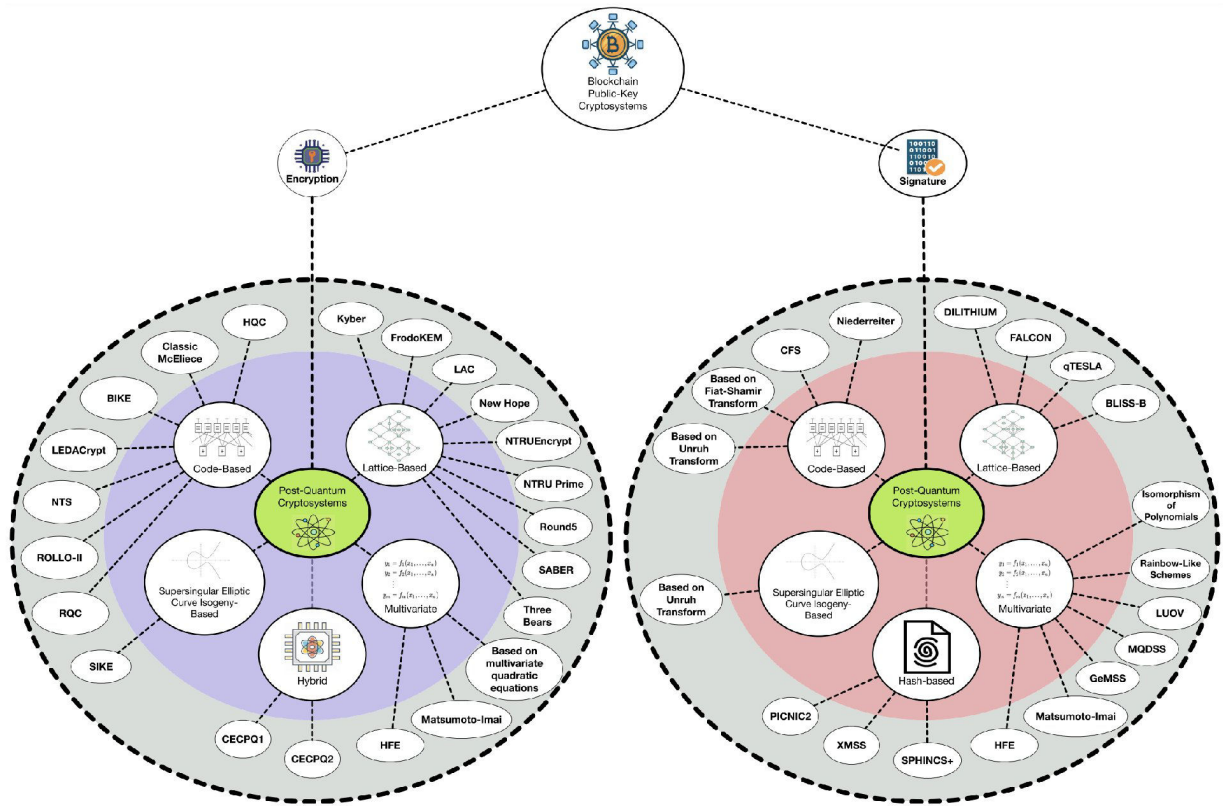

Fig. 3. Post-quantum public-key cryptosystem taxonomy and main practical implementations [16]

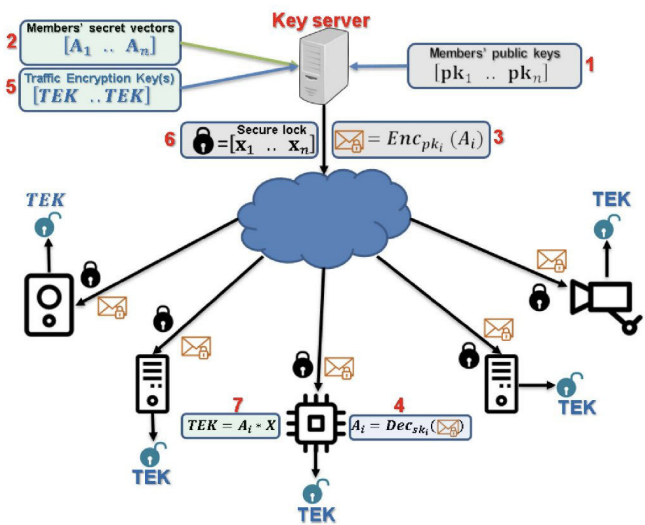

Fig. 4. ILLUSTRATES system model OF KeyShiled [19]

- Researchers DeMarcus Edwards and Danda B. Rawat [20] address the current status of quantum adversarial machine learning and proposed a novel technique by concentrating on the problems and proposed solutions. Problems in simulating quantum computers on classical computers remain unanswered; however, quantum machine learning algorithms can solve current problems in term of quantum probabilistic data-driven problems. The paper later defines the problems for quantum-assisted machine learning in near-term quantum for computers and limitations in datasets, applications, and adversarial examples. This paper commits that reader will get a clear understanding of quantum adversarial machine learning and helps research on this area.

- Quantum Computing companies, institutions and research groups may become targets by cybercriminals and hacktivists. To protect their networks, software, hardware and data from digital attacks Quantum applications become a necessity [4]. This paper describes the status quo of quantum computing technologies and the quantum threat associated with it. They define a threat vector for quantum computing systems and the respective defensive measures, mitigation and best practices to defend against threat,and recommend how to reduce the cyberattack by ensuring security by design of quantum software and hardware components.

TABLE IV
PRIMARY FINDINGS AND THEMES OF THE SELECTED STUDIES

| Selected Paper | Key Contribution | Types of Security | Year |
| --- | --- | --- | --- |
| [8] | Quantum random number generators for cybersecurity | Random number generators | 2021 |
| [16] | Introduces blockchain cryptography resistant to quantum computing attacks | Blockchains, DLTs & cryptography | 2020 |
| [18] | Post-quantum digital signature schemes based on using error-correcting codes | Cryptography and digital signature | 2020 |
| [19] | A scalable and quantum-safe key management scheme | Safe Key Management | 2020 |
| [21] | Quantum annealing based cybersecurity using restricted boltzmann machine | Quantum annealing | 2021 |
| [22] | Cybersecurity for quantum computing focuses on quantum threats | Quantum threats | 2018 |
| [23] | Quantum communication for post-pandemic cybersecurity | Cybersecurity in communication | 2022 |
| [24] | Quantum key distribution (MDI-QKD) based microgrid control architecture | Key distribution for cybersecurity | 2018 |
| [25] | Quantum cryptography for the future internet and the security | Quantum key distribution (QKD) | 2018 |
| [26] | Hybrid quantum-classical deep learning model for cybersecurity application | DGA-based botnet detection | 2021 |
| [27] | Authentication and encryption protocol based on quantum-inspired quantum walks (QIQW) | Authentication for cybersecurity | 2021 |
| [17] | Post-Quantum cybersecurity challenges associated with the Internet of Things | Post-quantum IoT security | 2018 |
| [16] | Post-Quantum Blockchain cryptography resistant to quantum computing attacks | Public-key cryptography and hash functions | 2021 |
| [28] | Quantum cryptography, quantum-key distribution (QKD) | Postquantum cryptography and cybersecurity | 2018 |
| [29] | Quantum computing by adopting restricted Boltzmann machine (RBM) | Bar-and-stripes (BAS) and cybersecurity | 2021 |
| [30] | Post-quantum security using CoreVUE breaks through PKI | Encryption and key exchange | 2022 |
| [31] | Cybersecurity education curricula through quantum computation | Cybersecurity education | 2021 |
| [32] | Traffic flow optimization using a quantum annealer | Traffic Flow using Quantum Computing | 2017 |
| [33] | Advanced encryption standard (AES) algorithm using quantum computing to encrypt/decrypt cybersecurity files | Encryption and decryption | 2021 |

- Quantum Computing companies, institutions, and research groups often become targets by cybercriminals and hacktivists. To protect their networks, software, hardware and data from digital attacks Quantum applications become a necessity [22]. The paper describes the status quo of quantum computing technologies and the quantum threat associated with it. Researchers define a threat vector for quantum computing systems and the respective defensive measures, mitigation, and best practices to defend against threats, and recommend how to reduce the cyberattack by ensuring security by design of quantum software and hardware components.

In another study, Hatma Suryotrisongko and Yasuo Musashi [26] proposed a novel hybrid quantum-classical deep learning model for cybersecurity applications that adopts domain generation algorithms (DGA) based botnet detection. The authors investigate the hybrid model's performance by comparing it with the classical model counterpart to investigate the quantum circuit's effectivity using four features of the botnet DGA dataset including MinREBotnets, CharLength, TreeNewFeature, and nGramReputation-Alexa. A combination of Pennylane's embedding (Angle Embedding and IQP Embedding) and layers circuit (Basic Entangler Layers, Random Layers, and Strongly Entangling Layers) makes The hybrid model's quantum circuit. The demonstration found hybrid model reached high performance (maximum accuracy up to 94,7% ) using their newly implemented noise models for assessing the applicability of the model and the combination of Angle Embedding and Strongly Entangled delivers high accuracy.

- In a different type of research, researcher Ronald P. Uhlig et al [4] discusses motivating graduate computer science and electrical engineering students in the study of quantum computing and cybersecurity. Teaching quantum computing at university-level computer science and electrical engineering programs is strategically important for workforce development with technological development has been appreciated by the U.S. Congress and other governmental and industrial organizations. Computer science and electrical engineering students don't want to study quantum computing because they need a clear understanding of physics, math, and computer science. The paper shows how to generate interest in quantum computing among graduate students and motivate them through exploring, in related courses using the scientific learning cycle as discussed by Wankat and oreoviczand introduce a student small group project in a cybersecurity course. The goal is to teach the students the basics of quantum computing by allowing them to"play with it" and let them know quantum computing is a growing new field that they need to follow. The project helped students to explore Public KeyCryptography which is more secure than current Public KeyCryptography. Such a strategy helps to motivate students to generate significant interest in quantum computing.

## IV. FINDINGS, DISCUSSION & NEW RESEARCH DIRECTION

The rapid growth of the internet and its popularization have led to the information age in human society, where people can use the internet globally. Cyberspace is used for communication, gaming, artistic media, and meetings, among others but is considered the largest unregulated platform in the world. Issues like computer viruses, hacking, and invasion by malicious software are the greatest threat to information security in computing. With these features of the quantum computer, the existing cryptography is threatened. New cryptosystems that are not based on discrete logarithms need to be explored to resist quantum computers. Quantum Information processing and communication are superior to classical communication due to their features. First, Uncertainty makes it impossible to determine the position of a particle in the microworld due to its existence in different places. Quantum cloning also makes it possible to retrieve any information deleted from the quantum by an attacker.

Quantum teleportation, additionally, allows the passing of data that has not been extracted in measurement by the sender to the receiver by measurement. Quantum information further possesses hidden features that are not possessed by classical information. Entanglement of the information in quantum cryptography makes it impossible to obtain the data through a local measurement operation; only joint measurement can expose it. Quantum cryptography is promising for the future of security in cyberspace. One of the desirable features of quantum cryptography is its unconditional security, guaranteed using cable and light in transmitting internet communication. The developments in quantum computing have additionally made it easier to solve the many complex problems in quantum physics. Unlike in classical computing, where data is protected through encryption, which is prone to eavesdropping by attackers, in quantum communication sniffing detection of the eavesdropper is easy due to the quantum no-cloning theory. Quantum cryptography is, therefore, more advantageous than classical cryptography due to its unconditional security feature and sniffing detection capabilities.

Although there are many challenges in quantum computing and accessibility in quantum devices, we study quantum machine learning for software supply chain attacks [34]. In a separate study, we also investigate the evolution of quantum computing and the use of quantum computing tools [35].

## V. CONCLUSION

The advent and development of quantum computing can be the most impactful domain for cybersecurity that can emerge both as a threat and solution to critical cybersecurity issues. In this study, we investigated the current state-of-the-art quantum computing and cybersecurity and presented the proposed approaches to date. Quantum computing for cyber security indicates that quantum computing shall be utilized for the betterment of cybersecurity threats. Besides, adopting quantum computing technology against cybersecurity poses threats. Our findings indicate the significant need for further research in this emerging domain. This timely systematic review shall provide a more thorough understanding and shed light on future research directions for quantum and cybersecurity practitioners and researchers.

## ACKNOWLEDGMENT

The work is partially supported by the U.S. National Science Foundation Awards 2100115, 1723578. Any opinions, findings, and conclusions or recommendations expressed in this material are those of the authors and do not necessarily reflect the views of the National Science Foundation.